\RequirePackage[T1]{fontenc}
\documentclass[conference]{IEEEtran}

\usepackage{amsmath,amsfonts}
\usepackage{comment} 
\usepackage{colortbl}
\usepackage{rotating}
\usepackage{booktabs,array,tabularx}

\usepackage{caption}
\usepackage[]{hyperref}
\usepackage{enumitem}
\usepackage{graphicx}

\usepackage{placeins} 
\usepackage{subcaption}
\usepackage[most]{tcolorbox}
\usepackage{textcomp}

\usepackage{wrapfig}
\usepackage{xcolor}
\usepackage{tikz}
\usepackage{pifont}
\usepackage{boldline}
\usepackage{multirow}
\usetikzlibrary{patterns}
\usepackage{float}

\usepackage{diagbox}
\usepackage[capitalise,nameinlink]{cleveref}

\def\BibTeX{{\rm B\kern-.05em{\sc i\kern-.025em b}\kern-.08em
    T\kern-.1667em\lower.7ex\hbox{E}\kern-.125emX}}

\newcommand{\fl}{
  \tikz\draw[black, thick] (0,0) circle (0.10cm);}

\newcommand{\pa}{
  \tikz\fill[black] (0,0) circle (0.11cm);}

\newcommand{\cm}[0]{\ding{51}}

\newcommand{\rcg}[0]{\rowcolor{gray!20}}

\usepackage[color=yellow]{todonotes}

\newcommand{\ransomware}[0]{OpenRansom}

\usepackage{ifthen}
\newboolean{fullversion}
\newcommand{\fullonly}[1]{\ifthenelse{\boolean{fullversion}}{#1}{}}
\newcommand{\shortonly}[1]{\ifthenelse{\boolean{fullversion}}{}{#1}}
\setboolean{fullversion}{false}

\begin{document}
\title{OpenRansom: A Configurable Lifecycle-Aware Ransomware Emulator}

\author{\IEEEauthorblockN{ 
Prateek Sahu\IEEEauthorrefmark{2},\;
Alexander Cathis\IEEEauthorrefmark{2},\;
Jessica Tseng\IEEEauthorrefmark{3},\;
Andreas Gerstlauer\IEEEauthorrefmark{2},\;
Mohit Tiwari\IEEEauthorrefmark{2}} 
\IEEEauthorblockA{\IEEEauthorrefmark{2}The University of Texas at Austin \;\; \IEEEauthorrefmark{3}IBM \;\; }
}


\maketitle
\begin{abstract}
Ransomware research needs repeatable experiments that expose more than the final encryption payload. A recovered binary may depend on unavailable infrastructure, while a fixed trace cannot reveal how behavior changes with target selection, exfiltration, or execution pacing. We present OpenRansom, a configurable emulator for studying selected post-compromise ransomware lifecycle stages. A study of 18 families from 2017--2024 informs its decomposition into reconnaissance, optional compression and exfiltration, defense-evasion controls, and encryption. Researchers can vary target scope, cryptographic choices, file-update behavior, and delays while collecting network, syscall, and hardware-counter observations. We describe the resulting stage-labeled dataset and its collection costs, including a reported average monitored-workload slowdown of 4.5\%. OpenRansom provides an experimental basis for studying attack variation; initial access, operational lateral movement, and detector design are outside this paper's contribution.

\end{abstract}
\section{Introduction}
Ransomware experiments must capture the activities that make encryption possible, as well as encryption itself. An attacker can enumerate mounted storage, identify valuable files, prepare archives, transfer data, and delay execution before denying access. Double extortion further separates the loss of confidentiality through exfiltration from the loss of availability through encryption~\cite{ransomhub,lockbit,revil}. A research workload that reproduces only the final payload cannot support controlled investigation of these earlier behaviors.

This creates an experimental problem distinct from designing a detector. Researchers need to choose which actions execute, vary their implementations, and observe the resulting behavior under a consistent measurement setup. File selection, transfer destinations, cipher choice, and execution pacing can all affect a defense's observations. When these dimensions are fixed inside a recovered sample, a result can be difficult to attribute to the behavior being studied rather than to a particular binary or environment.

Existing resources address parts of this problem. Malware repositories such as MalwareBazaar and VirusTotal provide access to samples~\cite{MalwareBazaar,VirusTotal}, but executing a binary does not guarantee that its supporting infrastructure is available or that all campaign stages can be reproduced. Public trace collections provide reusable observations~\cite{sgandurra2016automated,creech2013semantic,hirano2025ransmap,nkongolo2021ugransome1819,ispahany2026radar}, yet recorded traces cannot themselves be reconfigured to test a new target-selection or transfer policy. Emulation offers a complementary approach by exposing the choices that a researcher needs to control.

We present \ransomware{}, a configurable emulator organized around selected post-compromise ransomware behaviors. Its design is informed by the tools and techniques documented for 18 ransomware families from 2017--2024. Rather than treating a family name as an indivisible workload, OpenRansom decomposes behavior into selectable actions and parameters. The resulting configurations support local reconnaissance, optional preparation and transfer of data, execution delays and selected evasion controls, and encryption with configurable file-processing choices.

OpenRansom also connects configurable workloads to observation. We collect network traffic, OS system calls, and hardware performance counters during attack actions and benign execution. Stage-level labels allow downstream studies to compare signals for the same action or compose traces with different stage combinations. These labels describe the modeled experiment; they are not a claim that a synthetic trace reproduces every dependency or timing relationship of an operational attack.

Our contributions are:
\begin{enumerate}[nosep,leftmargin=*]
\item A lifecycle-oriented characterization of 18 ransomware families that motivates the emulator's configurable behavior and explicitly separates surveyed techniques from modeled capabilities.
\item A configurable post-compromise emulator with controls over target selection, exfiltration, encryption, file-update behavior, and execution delays.
\item A stage-labeled dataset spanning network, syscall, and hardware signals, together with measurements of collection latency, workload slowdown, and transient trace volume.
\end{enumerate}

This paper develops the emulation and data-generation of ransomware attack campaigns that can be subsequently used to develop and refine detection models. The goal of this project is to help synthesize ransomware signals into a reusable data format for further analysis.

\section{Background and Experimental Requirements}
\label{sec:background}
\subsection{Campaign Behavior and Experimental Scope}
Encryption-based extortion denies a victim access to selected data. In a double-extortion campaign, the attacker also retains a copy and can threaten disclosure. These objectives produce overlapping but separable activities: discovering data, preparing it for transfer, moving it to another destination, and encrypting the victim's copy~\cite{ransomhub,lockbit}. Their combination, rather than one cryptographic operation, motivates a lifecycle-oriented workload.

We organize the campaign context into initial access, persistence and command-and-control, reconnaissance and lateral movement, exfiltration, defense evasion, and execution. This organization is a descriptive model, not a requirement that every campaign execute every stage in a fixed order. Initial access establishes a foothold; persistence and command-and-control maintain access or deliver instructions. Reconnaissance identifies system state and candidate data. Exfiltration moves selected content away from the victim. Defense evasion can occur throughout the campaign, while execution applies the encryption payload.

OpenRansom starts after the foothold has been established. The broader lifecycle is useful for interpreting the family study, but the emulator is not a phishing platform, a persistence framework, or an implementation of operational multi-host compromise. Its purpose is to expose controllable system behaviors within the stated post-compromise scope.

\subsection{Behavioral Choices That Matter for Emulation}
Reconnaissance determines which directories, mounted volumes, and file types become targets. This matters experimentally because scanning a local hierarchy and accessing a mounted file share produce different system and network observations. Network discovery can be represented as a reconnaissance action without implementing subsequent lateral compromise. Credential theft and Active Directory exploitation remain outside the evaluated scope.

Exfiltration introduces both a destination and a preparation policy. A transfer can use a researcher-controlled endpoint or a configured cloud-storage destination, and compression can be included or omitted. Separating compression from transfer lets an experiment observe their individual costs and compose different stage sequences. The surveyed campaigns use several transfer utilities; support for every surveyed utility should not be inferred from that survey.

Encryption likewise admits multiple configurations. A hybrid scheme separates protection of key material from bulk file encryption. The symmetric algorithm, the fraction of each file processed, the write-and-delete ordering, and optional renaming affect execution behavior. OpenRansom exposes these choices rather than using a renamed file or a particular cipher as the definition of ransomware. Configurable sleep intervals provide an additional way to study sparse activity.

\subsection{Traces as an Experimental Artifact}
A binary sample, an emulator configuration, and a recorded trace answer different questions. A sample preserves a particular implementation; a configuration specifies modeled actions that can be rerun; a trace records observations from one execution environment. Our artifact combines configurable execution with stage-labeled traces. Network, syscall, and hardware measurements provide complementary views, but their presence does not by itself establish detector effectiveness.

\section{Ransomware Lifecycle Analysis}
\label{sec:lifecycle}
\begin{figure*}[h]
    \centering
    \includegraphics[trim={0.3cm 0.15cm 0.1cm 0.3cm},clip,width=\linewidth]{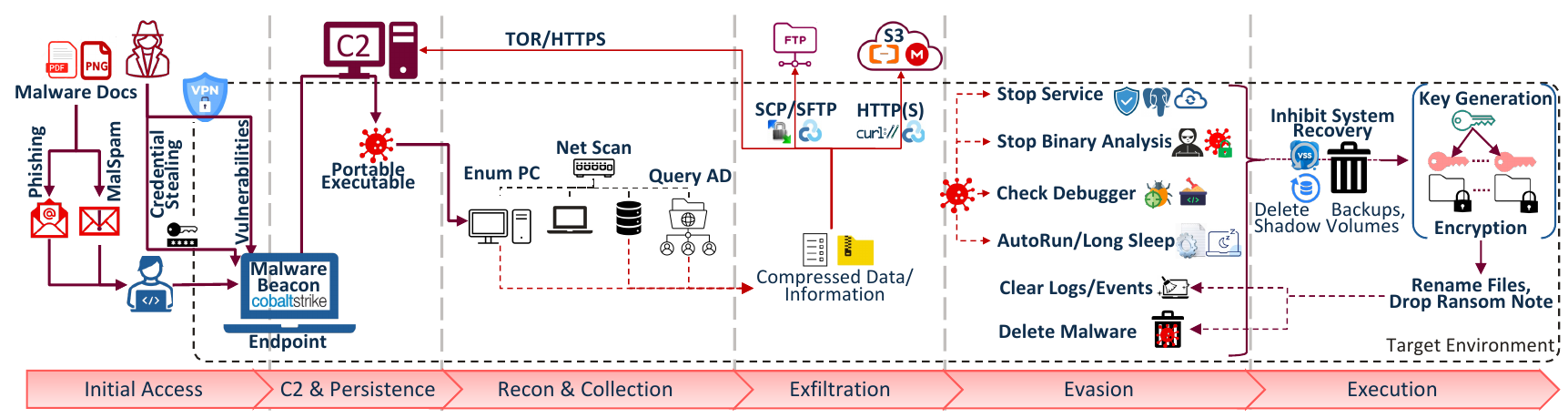}
    \caption{\small \bf Common tools and techniques used in real ransomware campaigns across their lifecycle stages.}
    \label{fig:lifecycle-flow}
\end{figure*}

We examine 18 ransomware families reported between 2017 and 2024. The family-level evidence is summarized in~\cref{tab:ransomware-tools}; \cref{fig:lifecycle-flow} organizes the common tools and techniques by lifecycle stage. These observations guide the selection of configurable actions rather than establish that OpenRansom implements each surveyed technique.

\newcommand{\rotc}[1]{\rotatebox[origin=c]{90}{\parbox{2cm}{\centering #1}}}
\begin{table*}[]

    \centering

    \resizebox{\textwidth}{!}{
    \begin{tabular}{V{3}cV{3}cV{3}cV{3}cV{3}c|c|cV{2}c|c|cV{3}cV{2}c|c|c|cV{2}cV{3}c|c|c|c|c|cV{3}cV{2}c|cV{2}c|cV{3}} \hline
        Ransomware & Year & Initial Access & Persistence & \multicolumn{6}{cV{3}}{Reconnaissance} & \multicolumn{6}{cV{3}}{Exfiltration} & \multicolumn{6}{cV{3}}{Defense Evasion} & \multicolumn{5}{cV{3}}{Execution} \\ \cline{1-4} \cline{5-27}
        &  &  &  & \rotc{Targeted File} & \rotc{Targeted Path} & \rotc{OS/Services} & \rotc{Active Directory} & \rotc{Memory Insp} & \rotc{Network Scan} & \rotc{Compress} & \rotc{rClone} & \rotc{cURL} & \rotc{SCP} & \rotc{Others} & \rotc{Location} & \rotc{Service/AV Stop} & \rotc{Modify Policies} & \rotc{Evade Analysis} & \rotc{Check Debugger} & \rotc{Long Sleep} & \rotc{Clear Logs} & \rotc{Inhibit Recovery} & \rotc{Asymmetric} & \rotc{Symmetric} & \rotc{Rename Files} & \rotc{Delete Binary} \\ \hline
        
        RansomHub~\cite{ransomhub}     & Aug '24 & Phishing   &             & \cm & \cm &     &     & \cm & \cm &     & \cm & \cm & \cm &     & S3      & \cm &     & \fl &     & \cm &     & \cm & x25519   & AES/ChaCha&     & \cm \\ \rcg
        Roadsweep~\cite{roadsweep}     & Jul '22 &            &             &     & \cm & \cm &     &     &     &    \multicolumn{6}{cV{3}}{No Exfil}   &     &     & \pa &     &     &     & \cm & RSA-2048 & RC4       & \cm & \cm \\
        LockBit~\cite{lockbit}         & Jun '22 &            & CobaltStrike&     &     & \cm & \cm & \cm & \cm & \cm & \cm &     & \cm & \cm & Mega    & \cm &     & \pa & \cm &     & \cm & \cm & RSA-2048 & AES/ChaCha& \cm & \cm \\ \rcg
        BlackBasta~\cite{blackbasta}   & Apr '22 & Phishing   &             & \cm & \cm & \cm &     &     &     &    \multicolumn{6}{cV{3}}{No Exfil}   &     &     &     & \cm &     &     & \cm & RSA-4096 & ChaCha    & \cm &     \\
        Cheerscrypt~\cite{cheerscrypt} & Jan '22 & Log4j      & CobaltStrike& \cm &     & \cm &     &     &     &     &     &     &     & \cm & Alibaba &     &     &     &     &     &     & \cm & ECDH     & Sosemanuk & \cm &     \\ \rcg
        BlackCat~\cite{alphv}          & Nov '21 & Cred Steal & CobaltStrike&     & \cm & \cm &     & \cm &     & \cm & \cm &     &     &     & Mega    & \cm &     &     &     &     & \cm & \cm &          &           & \cm &     \\
        HelloKitty~\cite{hellokitty}   & Jan '21 & SQL Inj.   & CobaltStrike&     & \cm & \cm & \cm &     &     & \cm & \cm &     &     & \cm & S3      &     &     &     &     &     &     & \cm & RSA-2048 & AES/Salsa & \cm &     \\ \rcg
        Babuk~\cite{babuk}             & Jan '21 & MalSpam    &             & \cm & \cm &     &     &     &     &     &     &     &     &     &         & \cm &     &     &     &     &     & \cm & NTRU     & ChaCha    & \cm & \cm \\
        Egregor~\cite{egregor}         & Sep '20 &            & Qakbot      & \cm & \cm & \cm & \cm &     &     & \cm & \cm &     &     &     & FTP Srv & \cm & \cm & \pa & \cm & \cm &     & \cm & RSA-2048 & AES/ChaCha& \cm &     \\ \rcg
        Conti~\cite{conti}             & Jul '20 & Phishing   &             &     & \cm &     & \cm & \cm &     &    \multicolumn{6}{cV{3}}{No Exfil}   & \cm &     & \pa &     &     &     & \cm & RSA-4096 & AES/ChaCha& \cm &     \\
        Avaddon~\cite{avaddon}         & Jun '20 &            &             & \cm &     & \cm &     &     &     &    \multicolumn{6}{cV{3}}{No Exfil}   &     &     &     &     &     &     & \cm & RSA-2048 & AES-256   &     &     \\ \rcg
        Netwalker~\cite{netwalker}     & May '20 & Phishing   &             &     &     &     &     & \cm &     &    \multicolumn{6}{cV{3}}{No Exfil}   & \cm &     & \pa &     &     &     & \cm & ECDH     &           & \cm &     \\
        REvil~\cite{revil}             & Mar '20 &            &             &     & \cm & \cm &     &     &     &     &     & \cm &     &     & FTP Srv & \cm &     & \pa &     &     &     & \cm & x25519   & AES/Salsa & \cm & \cm \\ \rcg
        ProLock~\cite{prolock}         & Mar '20 & Phishing   & Qakbot      &     &     &     & \cm &     &     &     & \cm &     &     &     & FTP Srv & \cm &     & \fl &     &     &     & \cm & RSA-2048 & RC6       &     & \cm \\
        Maze~\cite{maze}               & May '19 & MalSpam    &             & \cm & \cm & \cm & \cm & \cm & \cm & \cm & \cm &     & \cm &     & FTP Srv & \cm &     &     & \cm &     &     & \cm & RSA-2048 & ChaCha    & \cm &     \\ \rcg
        Cl0p~\cite{clop}               & Feb '19 & Phishing   &             & \cm &     &     &     &     &     &    \multicolumn{6}{cV{3}}{No Exfil}   & \cm & \cm & \fl &     &     &     & \cm & RSA-2048 &           & \cm &     \\
        Ryuk~\cite{ryuk}               & Sep '18 &            & Trickbot    &     & \cm & \cm &     &     &     &    \multicolumn{6}{cV{3}}{No Exfil}   & \cm &     & \fl &     &     & \cm & \cm & RSA-2048 & AES-256   & \cm &     \\ \rcg
        WannaCry~\cite{wannacry}       & May '17 &            &             & \cm &     & \cm &     &     &     &     &     &     &     & \cm & FTP Srv &     &     & \pa &     &     &     & \cm & RSA-2048 & AES-128   & \cm &     \\ \hline
        \ransomware  & -- & -- & -- & \cm & \cm & \cm &  &  & \cm & \cm & \cm &  & \cm &  & S3/FTP &  &  &  &  & \cm & \cm & \cm & Listed & Listed & \cm & \cm \\ \hline
    \end{tabular}
    }
    \caption{\small \bf An overview of tools and techniques used in a set of ransomware campaigns over the past decade. \cm represents specific tools used in corresponding ransomware. ``\protect\fl'' denotes that the malware was packed and obfuscated while ``\protect\pa'' denotes that these malware were packed, obfuscated and encrypted to avoid static binary analysis. }
    \label{tab:ransomware-tools}
\end{table*}

\paragraph{Reconnaissance.} The survey includes file and drive discovery, system and service enumeration, network discovery, memory inspection, and Active Directory exploration. OpenRansom uses local enumeration and configurable network discovery as workload actions; the presence of credential and directory-service techniques in the survey does not imply their implementation in the evaluated emulator.

\paragraph{Exfiltration.} The matrix records both campaigns with no observed exfiltration and campaigns using transfer utilities such as rClone, cURL, and SCP. Destinations include attacker-controlled servers and cloud services. This diversity motivates independent choices for compression and transfer destination. The described OpenRansom paths use rClone with S3/HTTPS or an SFTP endpoint; other surveyed tools remain campaign context.

\paragraph{Execution.} Families differ in their asymmetric and symmetric cryptography, file selection, renaming, and removal of their own binaries. OpenRansom therefore separates file-processing policy from cipher choice. The table's broad cryptographic capability labels summarize the options described in the emulator section; they do not mean support for every possible algorithm.

\paragraph{Defense evasion.} The surveyed reports include security-service termination, binary packing, analysis checks, sleep intervals, and log clearing. These are distinct capabilities. The emulator description includes configurable delays, selected software termination, and permissible log clearing, while the collected detector-evaluation dataset uses stage omission and interleaved sleep or benign execution to approximate evasive behavior. Packed binaries and debugger or sandbox checks are not established capabilities of that evaluated dataset.

\section{Threat Model and Scope}
\label{sec:ransom-tm}
We assume an attacker already has a foothold on the target endpoint and acts with the privileges available to the compromised user or process. OpenRansom models selected behaviors from that point: reconnaissance of local files and system state, optional network discovery, preparation and transfer of selected data, selected evasion controls, and encryption.

Initial access, phishing, payload delivery, persistent installation, and an operational command-and-control channel are outside the modeled scope. We also exclude privilege escalation, credential extraction, Active Directory exploitation, and lateral compromise of additional hosts. Access to a mounted NFS volume is an existing capability of the compromised endpoint; it does not constitute a modeled lateral-movement exploit. Researcher-controlled transfer endpoints provide destinations for the experiment, rather than reproduce attacker infrastructure.

The emulator is intended for controlled experiments on designated data. The source material describes configurable file paths, but does not establish a hard sandbox or a proof of containment. Reproducing a campaign's selected actions therefore remains distinct from claiming that the emulator reproduces the complete campaign or is intrinsically non-destructive.

\section{Emulator Design and Configuration}
\label{sec:emulator}
This section details the design of the \ransomware{} emulator and its configuration options, enabling practitioners to evaluate ransomware behavior in controlled environments.

\subsection{Modeled Stages and Configurations}
\paragraph{Reconnaissance:} The emulator enumerates the operating system, active services, and mounted drives, including network file systems and removable media. Target selection uses file extensions, modification dates, sizes, owners, and inclusion or exclusion lists. Configurable network discovery represents identification of connected systems or services. Although the surveyed campaigns include memory-based credential extraction and Active Directory scanning, those activities are outside the evaluated scope defined in~\cref{sec:ransom-tm}.

\paragraph{Exfiltration:} Data exfiltration has become an increasingly common tactic in ransomware campaigns to ensure ransom payment, particularly in cases of incomplete encryption, successful victim-side decryption, or as part of a \textit{double extortion} strategy. While not all ransomware families employ this technique, those that do typically use either cloud services or attacker-controlled FTP servers as the datastore, transferring files via HTTP(S) POST or SCP/SFTP protocols with tools such as rClone, WinSCP, or cURL. Our emulator models this behavior by allowing configurable destinations---either an S3 bucket over HTTPS or an SFTP server via rClone. Additionally, it supports optional compression of high-value targets into 7zip archives prior to transfer, simulating attacker efforts to accelerate uploads and reduce detection risk. 

\paragraph{Defense Evasion:} \ransomware{} models common detection evasion techniques, including terminating customizable list of target software, introducing configurable sleep intervals to evade statistical or behavioral detectors, and clearing event logs where permissible. 

\paragraph{Execution:} This stage runs the ransomware payload by encrypting the target files and data. Based on our study of prior ransomware campaigns, \ransomware{} provides many configurable options to ensure large coverage of different encryption techniques utilized as listed below, before placing a ransom note for the victim:
\begin{itemize}[leftmargin=*,nosep]
    \item Asymmetric Ciphers: RSA-4096, x25519, ECDH, NTRU
    \item Symmetric Ciphers: AES Block (ECB, CBC, CTR, CFB, OFB), Salsa20, ChaCha20
    \item Write-mode: In-place Overwrite, Write-after-Delete, or Delete-after-Write
    \item Rename: This renames encrypted files with a custom extension
    \item Encrypt-mode: Options to control the level of encryption on target files -- partial encryption (file headers or selected blocks), or complete encryption of file. 
    \item File path: Optional argument to limit ransomware to specified paths instead of entire system.
\end{itemize}

\section{Cross-Stack Data Collection}
\label{sec:datacollection}
The dataset captures three signal sources: network traffic, OS system calls (syscalls), and hardware performance counter (HPC) values. We collect network traffic and syscalls using \texttt{tshark} and \texttt{trace-cmd}, respectively, to capture behavior during ransomware and benign-workload execution. We sample HPCs using \texttt{perf} at 100\,ms intervals and extract average values for the \texttt{branch-instructions-retired}, \texttt{instructions}, \texttt{LLC-load-misses}, and \texttt{avx-insts.all} counters. 

We selected these counters by training a decision tree on 16 counters recommended in prior work~\cite{HASP24ransom,lifecycleMalwareAnalysis2014} and retaining the four with the highest feature importance across all lifecycle stages, as listed in~\cref{tab:key-features}. We segment ransomware samples into reconnaissance, compression, exfiltration, and encryption stages and benign workloads by application  in~\cref{tab:eval_phase_ransom}.
This dataset is open for analysis and development at \href{http://zenodo.org/records/18502469}{zenodo}.

\begin{table}[tb]
    \scriptsize
    \centering
    
    \resizebox{\columnwidth}{!}{%
    \begin{tabular}{|c||c|c|c|c|c|c|} \hline
        \cellcolor{white}Lifecycle & \multicolumn{6}{|c|}{Events ranked by Feature Importance (MDI)} \\ \cline{2-7}
        Stage & 1 & 2 & 3 & 4 & 5 & 6 \\ \hline \hline
        Recon & \textcolor{blue}{Inst} & \textcolor{blue}{Br Inst} & \textcolor{blue}{LLC Miss} & \textcolor{blue}{AVX Inst} & p2 uOps & Mem St \\ \hline
        Compress & \textcolor{blue}{Inst} & \textcolor{blue}{Br Inst} & p5 uOps & LLC Req & \textcolor{blue}{AVX Inst} & \textcolor{blue}{LLC Miss} \\ \hline
        Exfil & \textcolor{blue}{Inst} & \textcolor{blue}{LLC Miss} & \textcolor{blue}{Br Inst} & \textcolor{blue}{AVX Inst} & p2 uOps & LLC Req \\ \hline
        Encrypt & \textcolor{blue}{Inst} & \textcolor{blue}{LLC Miss} & \textcolor{blue}{Br Inst} & \textcolor{blue}{AVX Inst} & p2 uOps & Mem St \\ \hline
    \end{tabular}}
    \caption{\small \bf Top 6 distinguishing HPCs for each ransomware lifecycle stage, ranked by feature importance from a decision tree classifier trained on 16 HPCs. The 4 HPCs highlighted in blue were selected as the feature set used in evaluation.}
    \label{tab:key-features}
    
\end{table}

\newcommand{\rottext}[2]{\multirow{#1}{*}{\rotatebox[origin=c]{90}{{\centering #2}}}}
\begin{table}[]
\scriptsize
    \centering
    \resizebox{\columnwidth}{!}{
    \begin{tabular}{|c|p{5.5cm}|c|} \hline
        \textbf{Name} & \cellcolor{white} \textbf{Description} & \textbf{Stage}  \\ \hline\hline 
        Recon      & Explore target system, mount points and network  & RE \\ \cline{1-3}
        Compress   & Archive sensitive files -- gZip/zStd [Optional]  & F1 \\ \cline{1-3}
        Transfer   & Copy files/archives over network -- AWS S3/SFTP  & F2 \\ \cline{1-3}
                                
        AES        & Encrypt target file data using AES -- 128/256 bits & EX \\ \cline{1-3}
        Salsa      & Encrypt target file data using Salsa -- 128/256 bits & EX \\ \hline

    \end{tabular}
    }
    \caption{\small \bf Description of ransomware lifecycle stages, the variation in tool usage across stages in our evaluation.}
    
    \label{tab:eval_phase_ransom}
\end{table}

\section{Experimental Environment and Workloads}
The source experiments use CloudLab nodes with Intel Xeon E5-2660 v3 processors, Ubuntu 20.04, and Linux kernel 5.15. A compromised client, a researcher-controlled transfer/command endpoint, and an NFS server form the three-node testbed. The extra nodes supply storage and transfer services; they do not add a modeled lateral-compromise stage.

Attack measurements are segmented into reconnaissance, compression, transfer, and encryption actions, as listed in~\cref{tab:eval_phase_ransom}. The collected examples include AES and Salsa encryption, S3 and SFTP transfers, and optional compression. The wider configuration surface described in~\cref{sec:emulator} should be distinguished from this measured subset.

Benign measurements cover media serving with PhotoPrism~\cite{photoprism}, storage activity with Filebench~\cite{tarasov2016filebench}, Chromium workloads, and SPEC CPU 2017. These workloads exercise file traversal, bulk I/O, network transfer, and substantial computation. They provide useful comparison activity without assuming that encryption or file access alone implies malicious intent.

Stage-level samples also support downstream trace synthesis. In the Falcon study, feature vectors from attack actions and benign applications are concatenated with varied action durations, application changes, and omitted stages. Such sequences are constructed evaluation inputs, rather than recordings of complete live campaigns. This distinction matters when interpreting timing, causal dependencies, and the transferability of results to deployment.

\section{Collection Overhead}
We measure the collection pipeline at 100\,ms sampling intervals to characterize its suitability for online monitoring as well as offline trace generation.
We evaluate two overheads: system slowdown and resource usage.

\textit{Performance Overhead:}
We measure system slowdown as the percentage increase in execution time with active tracing. On average, we observe a 4.5\% increase across benign applications, driven by syscall tracing (2.11\%), network tracing (1.47\%), and HPC sampling (1.31\%). Syscall- and network-intensive workloads see higher, but acceptable overheads for large environments.

\begin{figure*}
    \centering
    \includegraphics[trim={0.3cm 0.3cm 0.25cm 0.75cm}, clip, width=0.75\textwidth]{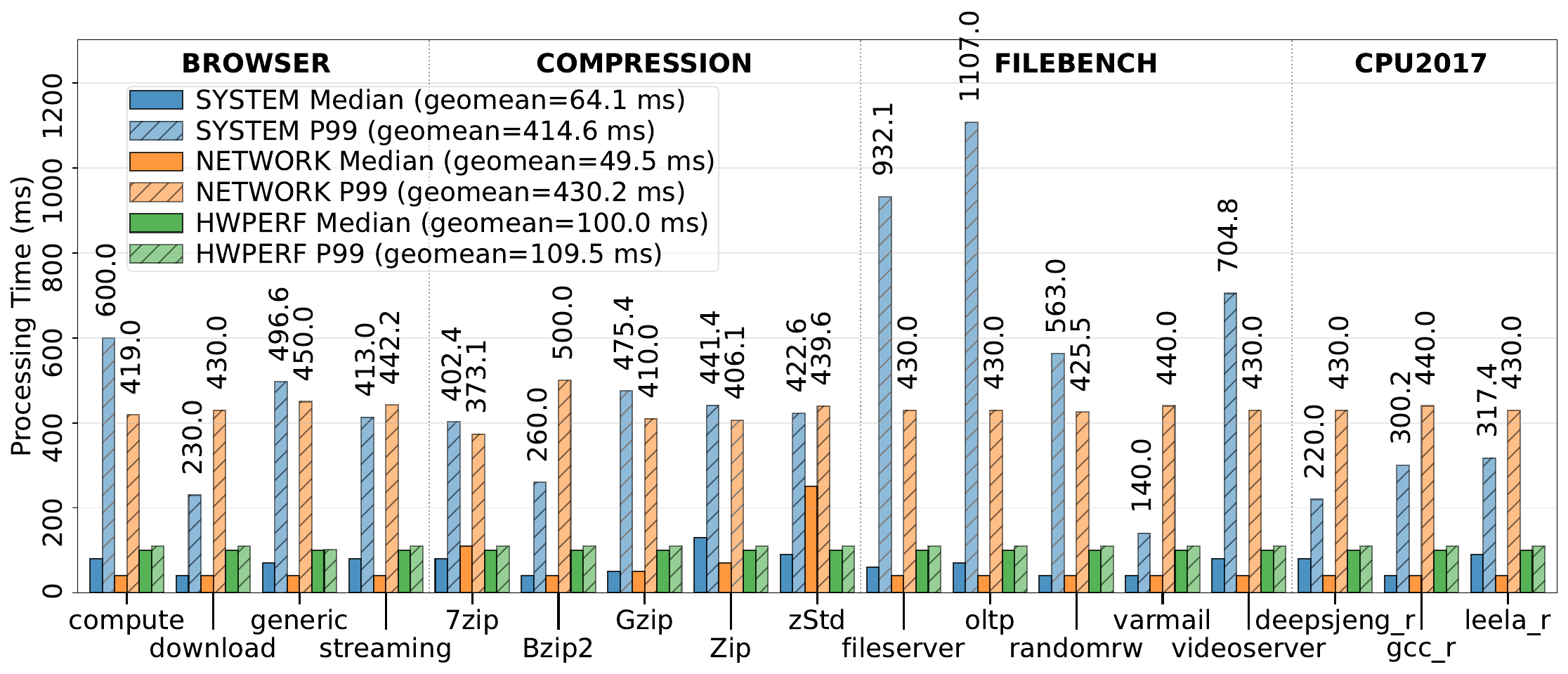}
    \caption{\small \bf Data processing latency by signal type at 10Hz. Median latencies remain under 100ms across all signal types, but p99 latencies vary significantly by workload, reaching up to 1.1s for syscalls and 500ms for network events.}
    \label{fig:processing_overhead}
\end{figure*}

\textit{Processing and Storage Overhead:}
For each 100ms sample, we collect a fixed number of HPC readings along with all observed syscall and network events, then process them for downstream analysis. \Cref{fig:processing_overhead} shows per-event processing latencies: median latency stays under 100ms across all signals and benchmarks, though p99 latency for syscalls reaches up to 1.1s and network events stay under 600ms. HPC samples are processed before the next collection window, while syscall and network processing may see 1-second backlogs under high activity.

Storage overhead similarly varies by workload (plot omitted for brevity): syscall tracing generates up to 6.73MB per sample for heavy workloads (geomean 811KB), network tracing adds up to 716KB per sample for high-throughput workloads (e.g., browser downloads, NFS accesses), and HPCs contribute a fixed 190 bytes per sample. Since data is processed and aggregated in real time rather than stored long-term, peak storage is transient, and subsampling, filtering, and compression can further reduce overhead.

\section{Discussion and Limitations}
\paragraph{Coverage and fidelity.} The family survey is a design basis, not a validation that every supported configuration faithfully reproduces an entire family. The evaluated workloads omit initial compromise, privilege escalation, Active Directory exploitation, and lateral compromise. A configuration captures selected behavior and may not preserve a real campaign's implementation details, dependencies, or timing.

\paragraph{Implementation versus measurement.} Configurable software termination and log clearing are described in the emulator design, but are not measured evasion behaviors in the detector dataset. Similarly, the stated cipher options exceed the AES and Salsa examples listed in the evaluated workload table. Further artifact-level validation is needed before treating the complete configuration space as experimentally covered.

\paragraph{Observation costs.} Median processing below the sampling interval does not establish a hard real-time guarantee: syscall tail latency reaches about 1.1\,s. Transient trace volume and collection-induced workload slowdown can also alter the behavior under study. Hardware features reflect the reported processor and toolchain, so portability and event availability need to be checked on other platforms.

\paragraph{Reproducibility.} The supplied materials identify an anonymized code repository and a dataset record. The dataset is linked in~\cref{sec:datacollection}; the code reference is \url{https://anonymous.4open.science/r/ransomware-setups-C694}. These are inherited artifact pointers, not a newly verified release. Future evaluation should report configuration manifests, repetitions, complete artifact inventories, and comparisons against held-out real samples where appropriate. Those details and experiments are not available in the supplied LaTeX sources.

\section{Related Work}
\paragraph{Samples and reusable observations.} MalwareBazaar and VirusTotal distribute samples that can ground studies in observed implementations~\cite{MalwareBazaar,VirusTotal}. Their role differs from exposing a controlled configuration interface. Public datasets provide reusable measurements, including system-call and ransomware-focused collections~\cite{creech2013semantic,sgandurra2016automated,hirano2025ransmap,nkongolo2021ugransome1819,ispahany2026radar}. OpenRansom complements recorded observations with configurable workload execution and measurements across three signal layers.

\paragraph{Emulators and simulators.} WannaLaugh and RanSim are relevant precedents for generating ransomware-like behavior~\cite{wannalaugh,RanSimK,ransim}. Our contribution is the explicit connection between a family-level lifecycle study, independently configurable post-compromise actions, and stage-labeled cross-stack measurements. The supplied source material does not contain a systematic head-to-head execution study of these tools, so we do not claim comparative fidelity or coverage based on new measurements.

\paragraph{Defensive systems.} File-system defenses such as Unveil, ShieldFS, and Redemption motivate the need for workloads that exercise more than one fixed file-encryption pattern~\cite{kharaz2016unveil,continella2016shieldfs,kharraz2017redemption}. OpenRansom supplies experimental inputs for defensive research but it does not contribute a classifier or recovery mechanism.

\section{Acknowledgments}
We thank IBM Security Research team members, especially Manoj Kumar, for the regular discussions and valuable feedback. 
This work was supported in part by ACE, one of the seven centers in JUMP 2.0, a Semiconductor Research Corporation (SRC) program sponsored by DARPA.

\section{Conclusion}
OpenRansom turns selected post-compromise ransomware behaviors into configurable experimental workloads. Its family study informs choices in reconnaissance, exfiltration, evasion controls, and encryption, while stage-labeled network, syscall, and hardware observations support downstream analysis. The reported collection costs characterize the measurement pipeline in the supplied testbed. Broader campaign fidelity, full configuration coverage, and portability remain subjects for further validation.

\bibliographystyle{plain}
\bibliography{content/references}

@misc{ransomhub,
  key = {ransomhub},
	title = {{RansomHub}, {Software} {S1212} {\textbar} {MITRE} {ATT}\&{CK}®},
	url = {https://attack.mitre.org/software/S1212/}
}

@misc{lockbit,
  key = {lockbit},
	title = {{LockBit} 3.0, {Software} {S1202} {\textbar} {MITRE} {ATT}\&{CK}®},
	url = {https://attack.mitre.org/software/S1202/}
}

@inproceedings{HASP24ransom,
author = {Woralert, Chutitep and Liu, Chen and Blasingame, Zander},
title = {Towards Effective Machine Learning Models for Ransomware Detection via Low-Level Hardware Information},
year = {2024},
isbn = {9798400712210},
publisher = {Association for Computing Machinery},
address = {New York, NY, USA},
url = {https://doi.org/10.1145/3696843.3696847},
doi = {10.1145/3696843.3696847},
booktitle = {Proceedings of the International Workshop on Hardware and Architectural Support for Security and Privacy 2024},
pages = {10--18},
numpages = {9},
location = {
},
series = {HASP '24}
}

@inproceedings{lifecycleMalwareAnalysis2014,
  author={Pandey, Sudhir Kumar and Mehtre, B.M.},
  booktitle={International Conference on Communication Systems and Network Technologies (CSNT)},
  title={A Lifecycle Based Approach for Malware Analysis},
  year={2014},
}

@misc{revil,
  key = {revil},
	title = {{REvil}, {Software} {S0496} {\textbar} {MITRE} {ATT}\&{CK}®},
	url = {https://attack.mitre.org/software/S0496/}
}

@misc{MalwareBazaar,
  key = {MalwareBazaar},
author = {},
  title = {MalwareBazaar | Malware sample exchange},
  url = "https://bazaar.abuse.ch/",
month = {},
year = {},
}

@misc{VirusTotal,
  key = {VirusTotal},
author = {},
  title = {VirusTotal - Home},
  url = "https://www.virustotal.com/gui/home/upload",
month = {},
year = {},
}

@article{sgandurra2016automated,
  title={Automated dynamic analysis of ransomware: Benefits, limitations and use for detection. arXiv 2016},
  author={Sgandurra, D and Mu{\~n}oz-Gonz{\'a}lez, L and Mohsen, R and Lupu, EC},
  journal={arXiv preprint arXiv:1609.03020},
  year={2016}
}

@article{creech2013semantic,
  title={A semantic approach to host-based intrusion detection systems using contiguousand discontiguous system call patterns},
  author={Creech, Gideon and Hu, Jiankun},
  journal={IEEE Transactions on Computers},
  volume={63},
  number={4},
  pages={807--819},
  year={2013},
  publisher={IEEE}
}

@article{hirano2025ransmap,
  title={RanSMAP: Open dataset of Ransomware Storage and Memory Access Patterns for creating deep learning based ransomware detectors},
  author={Hirano, Manabu and Kobayashi, Ryotaro},
  journal={Computers \& Security},
  volume={150},
  pages={104202},
  year={2025},
  publisher={Elsevier}
}

@article{nkongolo2021ugransome1819,
  title={Ugransome1819: A novel dataset for anomaly detection and zero-day threats},
  author={Nkongolo, Mike and Van Deventer, Jacobus Philippus and Kasongo, Sydney Mambwe},
  journal={Information},
  volume={12},
  number={10},
  pages={405},
  year={2021},
  publisher={MDPI}
}

@article{ispahany2026radar,
  title={RADAR: A realistic dataset for advancing ransomware detection},
  author={Ispahany, Jamil and Islam, Rafiqul and Deho, Oscar and Khan, M Arif and Islam, Zahid},
  journal={Journal of Cybersecurity},
  year={2026},
  publisher={Oxford University Press}
}

@misc{roadsweep,
  key = {roadsweep},
	title = {{ROADSWEEP}, {Software} {S1150} {\textbar} {MITRE} {ATT}\&{CK}®},
	url = {https://attack.mitre.org/software/S1150/}
}

@misc{blackbasta,
  key = {blackbasta},
	title = {Black {Basta}, {Software} {S1070} {\textbar} {MITRE} {ATT}\&{CK}®},
	url = {https://attack.mitre.org/software/S1070/}
}

@misc{cheerscrypt,
  key = {cheerscrypt},
	title = {Cheerscrypt, {Software} {S1096} {\textbar} {MITRE} {ATT}\&{CK}®},
	url = {https://attack.mitre.org/software/S1096/}
}

@misc{alphv,
  key = {alphv},
	title = {{BlackCat}, {Software} {S1068} {\textbar} {MITRE} {ATT}\&{CK}®},
	url = {https://attack.mitre.org/software/S1068/}
}

@misc{hellokitty,
  key = {hellokitty},
	title = {{HELLOKITTY}, {Software} {S0617} {\textbar} {MITRE} {ATT}\&{CK}®},
	url = {https://attack.mitre.org/software/S0617/}
}

@misc{babuk,
  key = {babuk},
	title = {Babuk, {Software} {S0638} {\textbar} {MITRE} {ATT}\&{CK}®},
	url = {https://attack.mitre.org/software/S0638/}
}

@misc{egregor,
  key = {egregor},
	title = {Egregor, {Software} {S0554} {\textbar} {MITRE} {ATT}\&{CK}®},
	url = {https://attack.mitre.org/software/S0554/}
}

@misc{conti,
  key = {conti},
	title = {Conti, {Software} {S0575} {\textbar} {MITRE} {ATT}\&{CK}®},
	url = {https://attack.mitre.org/software/S0575/}
}

@misc{avaddon,
  key = {avaddon},
	title = {Avaddon, {Software} {S0640} {\textbar} {MITRE} {ATT}\&{CK}®},
	url = {https://attack.mitre.org/software/S0640/}
}

@misc{netwalker,
  key = {netwalker},
	title = {Netwalker, {Software} {S0457} {\textbar} {MITRE} {ATT}\&{CK}®},
	url = {https://attack.mitre.org/software/S0457/}
}

@misc{prolock,
  key = {prolock},
	title = {{ProLock}, {Software} {S0654} {\textbar} {MITRE} {ATT}\&{CK}®},
	url = {https://attack.mitre.org/software/S0654/}
}

@misc{maze,
  key = {maze},
	title = {Maze, {Software} {S0449} {\textbar} {MITRE} {ATT}\&{CK}®},
	url = {https://attack.mitre.org/software/S0449/}
}

@misc{clop,
  key = {clop},
	title = {Clop, {Software} {S0611} {\textbar} {MITRE} {ATT}\&{CK}®},
	url = {https://attack.mitre.org/software/S0611/}
}

@misc{ryuk,
  key = {ryuk},
	title = {Ryuk, {Software} {S0446} {\textbar} {MITRE} {ATT}\&{CK}®},
	url = {https://attack.mitre.org/software/S0446/}
}

@misc{wannacry,
  key = {wannacry},
	title = {{WannaCry}, {Software} {S0366} {\textbar} {MITRE} {ATT}\&{CK}®},
	url = {https://attack.mitre.org/software/S0366/}
}

@inproceedings{wannalaugh,
author = {Diamantopoulos, Dionysios and Pletka, Roman and Sarafijanovic, Slavisa and Reddy, A.L. Narasimha and Pozidis, Haris},
title = {WannaLaugh: A Configurable Ransomware Emulator - Learning to Mimic Malicious Storage Traces},
year = {2024},
isbn = {9798400711817},
publisher = {Association for Computing Machinery},
address = {New York, NY, USA},
url = {https://doi.org/10.1145/3688351.3689163},
doi = {10.1145/3688351.3689163},
booktitle = {Proceedings of the 17th ACM International Systems and Storage Conference},
pages = {118--131},
numpages = {14},
location = {Virtual, Israel},
series = {SYSTOR '24}
}

@misc{RanSimK,
  key = {RanSimK},
author = {},
  title = {RanSim | KnowBe4},
  url = "https://www.knowbe4.com/free-cybersecurity-tools/ransim",
month = {},
year = {}
}

@misc{ransim,
  key = {ransim},
author = {},
  title = {lawndoc/RanSim: Ransomware simulation script written in PowerShell. Useful for testing your defenses and backups against real ransomware-like activity in a controlled setting.},
  url = "https://github.com/lawndoc/RanSim/tree/main",
month = {},
year = {},
}

@inproceedings{kharaz2016unveil,
  title={$\{$UNVEIL$\}$: A $\{$Large-Scale$\}$, automated approach to detecting ransomware},
  author={Kharaz, Amin and Arshad, Sajjad and Mulliner, Collin and Robertson, William and Kirda, Engin},
  booktitle={25th USENIX security symposium (USENIX Security 16)},
  pages={757--772},
  year={2016}
}

@inproceedings{continella2016shieldfs,
author = {Continella, Andrea and Guagnelli, Alessandro and Zingaro, Giovanni and De Pasquale, Giulio and Barenghi, Alessandro and Zanero, Stefano and Maggi, Federico},
title = {ShieldFS: a self-healing, ransomware-aware filesystem},
year = {2016},
isbn = {9781450347716},
publisher = {Association for Computing Machinery},
address = {New York, NY, USA},
url = {https://doi.org/10.1145/2991079.2991110},
doi = {10.1145/2991079.2991110},
booktitle = {Proceedings of the 32nd Annual Conference on Computer Security Applications},
pages = {336--347},
numpages = {12},
location = {Los Angeles, California, USA},
series = {ACSAC '16}
}

@inproceedings{kharraz2017redemption,
  title={Redemption: Real-time protection against ransomware at end-hosts},
  author={Kharraz, Amin and Kirda, Engin},
  booktitle={International Symposium on Research in Attacks, Intrusions, and Defenses},
  pages={98--119},
  year={2017},
  organization={Springer}
}

@misc{photoprism,
  key = {photoprism},
author = {},
  title = {photoprism/photoprism: AI-Powered Photos App for the Decentralized Web},
  url = "https://github.com/photoprism/photoprism?tab=readme-ov-file",
month = {},
year = {}
}

@article{tarasov2016filebench,
  title={Filebench: A flexible framework for file system benchmarking},
  author={Tarasov, Vasily and Zadok, Erez and Shepler, Spencer},
  journal={USENIX; login},
  volume={41},
  number={1},
  pages={6--12},
  year={2016}
}
\end{document}